\documentclass[11pt]{article}
\usepackage{amsthm}
\usepackage{amsmath}
\usepackage{amssymb}
\usepackage{mathdots}
\usepackage{esint}
\usepackage{authblk}

\makeatletter
\theoremstyle{plain}
\newtheorem{thm}{\protect\theoremname}
  \theoremstyle{plain}
  
  \theoremstyle{definition}
  \newtheorem{example}[thm]{\protect\examplename}

\makeatother

  \providecommand{\examplename}{Example}
  \providecommand{\lemmaname}{Lemma}
\providecommand{\theoremname}{Theorem}

\begin{document}

\title{On Functionally Commutative Quantum Systems}

\author{Takeo Kamizawa}
\affil{{\footnotesize Faculty of Physics, Astronomy and Informatics, Nicolaus Copernicus University, Toru{\'n}, Poland}}
\maketitle
\begin{abstract}
In this paper,  a method to solve functionally commutative time-dependent linear homogeneous differential equation is discussed. We apply this technique to solve some dynamical quantum problems. 
\end{abstract}

\section{Introduction}

The study of natural phenomena and their model analysis often
involves differential equations, and it is well-known that once the
initial condition is stated there is the unique solution for a linear differential
equation. If we could obtain the solution of a differential equation
in a closed form, we would be able to follow a trajectory completely, i.e.
we can analyse the system well. However, the problem is that the class
of integrable (or solvable) differential equations using the methods
known today is quite small, and we do not have a uniform technique to solve
for all differential equations. Thus, we look for the closed forms of the solutions of
some particular types of differential equations using specific methods. 

The dynamics in quantum mechanics is expressed by linear equations
on the set of all bounded operators $\mathcal{B}\left(\mathcal{H}\right)$ on a Hilbert space $\mathcal{H}$:
\[
\frac{d}{dt}\rho\left(t\right)=L\left(t\right)\rho\left(t\right),
\]
but the general closed solution is not known if $L$ is time-dependent. One
of the well-studied class is the one such that the generator of the time evolution $L\left(t\right)$
is functionally commutative, i.e. $L\left(t\right)L\left(s\right)=L\left(s\right)L\left(t\right)$
for $t,s\in\mathcal{I}$ on some open interval $\mathcal{I}$, and this
class of open quantum systems is analysed in \cite{chrusinski2010,chrusinski2015},
where the solution of functionally commutative quantum systems were obtained using
the spectrum of $L\left(t\right)$. This technique is useful when
the dimension is small (e.g. $\dim\mathcal{H}=1,2,3$) because the
spectrum can be calculated relatively easily, but it becomes difficult
when the dimension increases, especially when $\dim\mathcal{H}\geq5$
because of the Galois theory. 

However, linear systems with functionally commutative generators have been studied and there are many important papers in this field  \cite{ascoli1950,bogdanov1959,epstein1962,rose1965,erugin1966,martin1967,goff1981,kotin1982,evard1985,zhu1992,holtz2013}. Especially, Zhu's technique \cite{zhu1990,zhu1992} enables us to compute the solutions of these differential equations effectively. 
A benefit of this method is that this technique does not rely on the calculation of the eigenvalues of $L\left(t\right)$, so we may be able to obtain the solutions of high-dimensional differential equations, where the spectrum of the generator is seldom computable.

In this paper, we study functionally commutative differential equations. We study Zhu's technique in Sect. 2, and as examples this technique is applied in Sect. 3 to some 2-level quantum systems with commutative generators.

\section{Preliminaries}

\subsection{Open Quantum Systems}

First of all, let us introduce several notations and concepts in open quantum
systems (for details, see \cite{nielsen2010,jamiolkowski2013,jamiolkowski1992}). Let $\mathcal{H}_{S}$ be an $n$-dimensional Hilbert space corresponding to a quantum system $S$,
$\mathcal{L}\left(\mathcal{H}_{S}\right)$ be the set of all linear
transformations on $\mathcal{H}_{S}$ and $\mathcal{B}_{*}\left(\mathcal{H}_{S}\right)\subset\mathcal{L}\left(\mathcal{H}_{S}\right)$
be the set of all self-adjoint bounded operators. The set of states
on $\mathcal{H}_{S}$ is defined by 
\[
\mathcal{S}\left(\mathcal{H}_{S}\right)=\left\{ \rho\in\mathcal{B}_{*}\left(\mathcal{H}_{S}\right)\mid\rho\geq0,\;\mathrm{tr}\rho =1\right\} ,
\]
where $\rho\geq0$ means that $\rho$ is positive semidefinite, i.e.
$\left\langle \varphi\right|\rho\left|\varphi\right\rangle \geq0$
for all $\left|\varphi\right\rangle \in\mathcal{H}_{S}$. The dynamics
on $\mathcal{S}\left(\mathcal{H}_{S}\right)$ is represented by a
differential equation called a master equation as
\begin{equation}
\frac{d}{dt}\rho\left(t\right)=L\left(t\right)\rho\left(t\right),\label{eq:master_equation}
\end{equation}
where $L\left(t\right):\mathcal{L}\left(\mathcal{H}_{S}\right)\to\mathcal{L}\left(\mathcal{H}_{S}\right)$
is a linear operator (often called a generator) on an open interval
$\mathcal{I}\subset\mathbb{R}$ containing 0. The solution of (\ref{eq:master_equation}) with
the initial state $\rho_{0}\in\mathcal{S}\left(\mathcal{H}_{S}\right)$
is a transformation $\rho:\mathcal{I} \to\mathcal{S}\left(\mathcal{H}_{S}\right)$
which satisfies (\ref{eq:master_equation}) and $\rho\left(0\right)=\rho_{0}$,
and it is known that $\rho\left(t\right)$ for (\ref{eq:master_equation})
is uniquely determined (cf. e.g. \cite{nielsen2010,lukes1982}) and the
solution is expressed using the linear transformation $\Phi:\mathcal{I}\times\mathcal{S}\left(\mathcal{H}_{S}\right)\to\mathcal{S}\left(\mathcal{H}_{S}\right)$
called the fundamental solution (or dynamical map, especially in quantum physics) as
\begin{equation}
\rho\left(t\right)=\Phi\left(t\right)\rho_{0}.\label{eq:solution_master_equation}
\end{equation}
By putting (\ref{eq:solution_master_equation}) into (\ref{eq:master_equation}),
we also obtain the equation of this fundamental solution: 
\begin{equation}
\frac{d}{dt}\Phi\left(t\right)=L\left(t\right)\Phi\left(t\right)\label{eq:master_equation_operator}
\end{equation}
with $\Phi\left(0\right)=I$ (identity). 

Geometrically, the set of states $\mathcal{S}\left(\mathcal{H}_{S}\right)$
forms a convex set, so $\Phi$ has to make this set invariant during its evolution, i.e. $\Phi$ maps from a positive element to a positive element (such a map is called a positive map). For making $\Phi$ a positive map, it is known that $L$ has to be an operator called a Metzler
operator and this is also a sufficient condition (see \cite{jamiolkowski2013,dragan2005}). 

In the study of open quantum systems, we assume that our system $\left(\mathcal{H}_{S},\Phi\right)$
is affected by its surroundings or environment, which is also supposed
to form another system $\left(\mathcal{H}_{E},\Gamma\right)$ on $\mathcal{I}$,
and the system $\left(\mathcal{H}_{S}\otimes\mathcal{H}_{E},\Phi\otimes\Gamma\right)$
is assumed to form the total system on $\mathcal{I}$. Indeed, the
total system is considered to form a closed system, namely a system
such that 
\begin{equation}
L\left(t\right)\rho  =  -\frac{i}{\hbar}\left[H\left(t\right),\rho\right],\label{eq:vonNeumann_equation}
\end{equation}
where $\hbar>0$ is some constant, $[A,B]=AB-BA$ is the commutator, $H\left(t\right):\mathcal{H}_{S}\otimes\mathcal{H}_{E}\to\mathcal{H}_{S}\otimes\mathcal{H}_{E}$
is some time-dependent self-adjoint linear operator called a Hamiltonian,
and the equation (\ref{eq:vonNeumann_equation}) is called the
von Neumann equation. 

For the analysis of our system $(\mathcal{H}_{S},\Phi)$, we are interested in the closed form of the solution $\Phi$ so that we will be able to follow all orbits in the system. However, because of the noise from the environment the structure of $L(t)$ can be complicated and it may be difficult to obtain the closed form of $\Phi$. 
In this field, some models have been considered and analysed, and one of them is the functionally commutative class, where $\left[ L(t),L(s) \right]=O$ for $t,s\in \mathcal{I}$ \cite{chrusinski2010,chrusinski2015}. The main topic in this paper is about open quantum systems of this class, and we discuss the integrability later.

\subsection{Magnus Expansion}

For a linear differential equation (\ref{eq:master_equation}) with the variable coefficient $L(t)$, we are interested in the unique solution of this equation. If the generator is time-independent, i.e. $L(t)=L$ for all $t\in \mathcal{I}$, the solution is given by 
\begin{equation}
\rho(t)=\exp \left( Lt \right) \rho_{0} \label{eq:time-independent_solution_exponential}
\end{equation}
for a given initial state $\rho_{0}\in \mathcal{S}\left( \mathcal{H}_{S} \right)$, where
\[ \exp(Lt) =\sum_{k=0}^{\infty} \frac{t^{k}}{k!} L^{k} = \sum_{k=0}^{m-1} \alpha_{k}(t) L^{k}, \]
$m$ is the degree of the minimal polynomial of $L$ and $\alpha_{k}:\mathcal{I} \to \mathbb{C}$ are some scalar functions. 

We may expect that a similar theory should hold for a time-dependent generator $L(t)$, but in fact the solution of (\ref{eq:master_equation}) with time-dependent generator $L(t)$ cannot be represented like (\ref{eq:time-independent_solution_exponential}) in general.

Magnus \cite{magnus1954} gave a sufficient condition for this problem:
\begin{thm}
\cite[Theorem 1]{blanes2009}Let $\Phi\left(t\right)$ be a map satisfying
(\ref{eq:master_equation_operator}) and $\Phi\left(0\right)=I$.
If 
\[ \frac{d}{dt}\Phi\left(t\right)=\sum_{k=0}^{\infty}\frac{B_{k}}{k!}\mathrm{ad}_{\Phi\left(t\right)}^{k}L\left(t\right) \]
converges for all $t$, where $B_{k}$ is the $k$-th Bernoulli number
and 
\[
\mathrm{ad}_{A}^{0}B=B,\;\mathrm{ad}_{A}^{1}B=\left[A,B\right],\;\mathrm{ad}_{A}^{k}B=\left[A,\mathrm{ad}_{A}^{k-1}B\right],
\]
then $\Phi\left(t\right)$ is given by 
\[
\Phi\left(t\right)=\exp \left(  L\left(t \right) \right).
\]
Moreover, $\Phi(t)$ has an infinite series representation, which is now usually called the Magnus expansion:
\begin{equation}
\Phi(t)=\int_{0}^{t} L(t_{1}) dt_{1} -\frac{1}{2} \int_{0}^{t} \left[ \int_{0}^{t_{1}} L(t_{2})dt_{2},L(t_{1}) \right] dt_{1}+\cdots. \label{eq:magnus_expansion}
\end{equation}

\end{thm}
The Magnus expansion is a good formulation
for the solution of a general linear differential equation of the
form (\ref{eq:master_equation}) with the time-dependent generator.
However, the Magnus expansion is an infinite series, which is difficult to be dealt with, so we are next interested in linear systems which has a finite series expression, which is discussed in the following sections.

\subsection{Linear Differential Equation of Functionally Commutative Systems}

In this section we study a particular type of differential equations; differential equations of the form (\ref{eq:master_equation}) with the generators $L(t)$ which commute with its derivative on $\mathcal{I}$, i.e. 
\begin{equation}
\left[ L(t), \frac{d}{dt}L(t) \right]=O. \; \left( \forall t\in \mathcal{I} \right) \label{commutingderivative}
\end{equation}

The study of this type of differential equations seems to have started before 1934 \cite{holtz2013}, and many discussions have been done on this problem \cite{ascoli1950,bogdanov1959,epstein1962,rose1965,erugin1966,martin1967,goff1981,kotin1982,evard1985,zhu1992,holtz2013}. 

One reason of the popularity of this type of systems is that the solution of (\ref{eq:master_equation}) satisfying (\ref{commutingderivative}) is known to be \cite{evard1985,lukes1982}: 
\begin{equation}
\rho(t)=\exp \left( \int_{0}^{t} L(\tau) d\tau \right) \rho_{0}. \label{eq:solution_initial_value_problem}
\end{equation}
In addition, in some cases the generator has a finite-sum form 
\[ L(t)=\sum_{k=0}^{M} \alpha_{k}(t) L_{k}, \]
where $L_{k}$ are linearly independent constant matrices and $\alpha_{k}$ are scalar functions \cite{ascoli1950,kotin1982,evard1985}, where we are able to calculate the closed form
\[ \Phi(t)=\exp \left( \int_{0}^{t} L(\tau) d\tau \right) \] easily. 

It is important to remark that the condition (\ref{commutingderivative}) is equivalent to the commutativity of the generator $L(t)$ because of the following theorem:
\begin{thm}
\cite[Theorem 7.4.1]{lukes1982}. For integrable function $A(t)$ on $\mathcal{I}$ the following conditions are equivalent:
\begin{itemize}
\item $\left[ A(t),A(s) \right]=O$ for all $t,s\in \mathcal{I}$;
\item $\left[ A(t), \int_{s}^{t} A(\tau) d\tau \right]=O$ for all $t,s\in \mathcal{I}$.
\end{itemize}
\end{thm}
Putting $A(t)=\frac{d}{dt} L(t)$, we obtain the equivalence of the functional commutativity and the commutivity of the generator with its derivative. Therefore, the functionally commutative systems have some ideal properties like the systems of the form (\ref{commutingderivative}). 

\subsection{Finite Form of the Exponential}

Magnus expansion (\ref{eq:magnus_expansion}) is a useful formula to calculate the closed form of the solution $\Phi(t)$ for some systems.  
However, the Magnus expansion involves an infinite series, which is
difficult to be dealt with. For applications and computational purposes,
it is beneficial for us if the series can be also represented by a
finite series. 

The key starting point for this problem would be the result by Martin \cite[Theorem 2]{martin1967}:
\begin{thm}
Suppose $L(t)$ is bounded and piecewise continuous for $t\geq 0$. Then $L(t)$ is functionally commutative on $t\geq 0$ iff there is a set of mutually commuting constant matrices $\{ L_{k}=L(t_{k}) \}_{k=1}^{M}$ for some $M\leq n^{2}$ and bounded piecewise continuous scalar functions $\alpha_{k}$ such that 
\[ L(t)=\sum_{k=1}^{M} \alpha_{k}(t) L_{k}. \]
\end{thm}

A benefit of this theorem is that we are able to simply calculate 
\[ \Phi(t)=\exp \left( \int_{0}^{t} L(\tau)d\tau \right) \]
using the decomposition by Martin. However, as Zhu pointed out in \cite{zhu1990}, there are difficulties in finding the decomposition. In \cite{zhu1990,zhu1992} Zhu constructed a systematic method to obtain a finite decomposition of a functionally commutative generator $L(t)$, called a spatial decomposition. In this section his method and the application to calculate $\exp \left( \int_{0}^{t} L(\tau) d\tau \right)$ are introduced. 

Let $\mathbb{L}=\mathbb{L}\left(\mathcal{I},\mathbb{C}\right)$ be
a linear space of maps $f:\mathcal{I}\to\mathbb{C}$. A matrix function
$F\in\mathcal{M}_{n}\left(\mathbb{L}\right)$ is said to be proper
if there are $G\in\mathcal{M}_{n}\left(\mathbb{C}\right)$, $p\in\mathbb{N}$
and $\alpha_{k}\in\mathbb{L}$ ($k=1,\ldots,p$) such that 
\[
F\left(t\right)=\sum_{k=1}^{p}\alpha_{k}\left(t\right)G^{k-1},
\]
and the associated function $f:\mathcal{I}\times\mathbb{C}\to\mathbb{C}$:
\[
f\left(t,\lambda\right)=\sum_{k=1}^{p}\alpha_{k}\left(t\right)\lambda^{k-1}
\]
is called a primitive function. Note that the primitive function need
not be defined on the entire complex plane but is required that, for
some $D\subset\mathbb{C}$, $f\left(\cdot,\lambda\right)\in\mathbb{L}$
for all $\lambda\in D$, $\lambda_{i}\in D$ and 
\[
\left\{ \left.\frac{1}{k_{i}!}\cdot\frac{\partial^{k_{i}}f\left(t,\lambda\right)}{\partial\lambda^{k_{i}}}\right|_{\lambda=\lambda_{i}}\mid k_{i}=0,1,\ldots,m_{i}-1,\; i=1,2,\ldots,r\right\} \subset\mathbb{L},
\]
where $\sigma\left(G\right)=\left\{ \left(\lambda_{i},m_{i}\right)\right\} _{i=1}^{r}$
is the spectrum (i.e. $\lambda_{i}$ is an eigenvalue and $m_{i}$
is its multiplication) of $G$. 

Recall that a matrix function $F\left(t\right)$ is said to be functionally commutative
(or semiproper) on $\mathcal{I}$ if 
\[
F\left(t\right)F\left(s\right)=F\left(s\right)F\left(t\right)\;\left(\forall t,s\in\mathcal{I}\right).
\]
An important discovery by Zhu was that a functionally commutative matrix function
$F$ can be decomposed into a finite sum of proper functions as follows:
\begin{thm}
$F\in\mathcal{M}_{n}\left(\mathbb{L}\right)$ is functionally commutative on $\mathcal{I}$
iff $F$ can be decomposed into a sum of proper functions (called
a spatial decomposition): 
\[
F\left(t\right)=\sum_{i=1}^{m}F_{i}\left(t\right)=\sum_{i=1}^{m}f_{i}\left(t,G_{i}\right),
\]
where $\left[G_{i},G_{j}\right]=O$ for all $i,j\leq m$. 
\end{thm}
Using the spatial decomposition of a functionally commutative matrix function,
we are able to solve the linear differential equation (\ref{eq:master_equation}) according to the following theorem: 
\begin{thm}
\label{thm:zhu_spatial_decomposition_solution}\cite[Theorem 4]{zhu1992}. If
$L\in\mathcal{M}_{n}\left(\mathbb{K}\right)$ is semiproper on $\mathcal{I}$
having a decomposition 
\[
L\left(t\right)=\sum_{i=1}^{m}L_{i}\left(t\right)=\sum_{i=1}^{m}f_{i}\left(t,G_{i}\right),
\]
where $\left[G_{i},G_{j}\right]=0$ and $m$ is the degree of the
minimal polynomial of $L$, then the system (\ref{eq:master_equation})
has the finite solution 
\[
\Phi\left(t\right)=\prod_{i=1}^{m}g_{i}\left(t,G_{i}\right),
\]
where 
\[
g_{i}\left(t,\lambda\right)=\exp\left(\int_{0}^{t}f_{i}\left(\tau,\lambda\right)d\tau\right).
\]

\end{thm}

\section{Functionally Commutative Quantum Systems}

As shown in the previous section, a functionally commutative linear homogeneous differential equation can be solved using the technique by Zhu. In this section we apply this method to some functionally commutative quantum systems. 

\begin{example}
In \cite{chrusinski2015}, a functionally commutative system:
\begin{equation}
L(t)\rho =\sum_{k=1}^{n^{2}-1} \alpha_{k}(t) \left[ U_{k}\rho U_{k}^{*} -\rho \right] \label{commexample1}
\end{equation}
was analysed, where $U_{k}$ are unitary operators such that $\{ U_{k} \}_{k} \cup \{ I \}$ forms a basis of $\mathcal{L}(\mathcal{H})$. In this case, the eigenvalues of $L$ can be obtained relatively easily, and the solution was finally calculated using the spectrum. 

Alternatively, as an example, we calculate the solution of (\ref{commexample1}) with $n=2$ \cite[Example 1]{chrusinski2015} using the Zhu's technique. Let
\[
\frac{d}{dt}\rho_{t}=L\left(t\right)\rho_{t}=\gamma\sum_{k=1}^{3}\alpha_{k}\left(t\right)\left(\sigma_{k}\rho_{t}\sigma_{k}^{*}-\rho_{t}\right),
\]
where $\gamma\in\mathbb{C}$ is some constant, $\alpha_{k}:\mathbb{R}\to\mathbb{C}$
are some complex functions and $\sigma_{k}$ are the Pauli matrices:
\[
\begin{array}{ccc}
\sigma_{1}=\left(\begin{array}{cc}
0 & 1\\
1 & 0
\end{array}\right), & \sigma_{2}=\left(\begin{array}{cc}
0 & i\\
-i & 0
\end{array}\right), & \sigma_{3}=\left(\begin{array}{cc}
1 & 0\\
0 & -1
\end{array}\right).\end{array}
\]
According to the
identity \cite[pp.35]{magnus1999}: 
\[
\mathrm{vec}\left(ABC\right)=\left(C^{T}\otimes A\right)\mathrm{vec}B,
\]
the matrix form of $L\left(t\right)$ is given
by 
\begin{eqnarray}
\tilde{L}\left(t\right) & = & \gamma\sum_{k=1}^{3}\alpha_{k}\left(t\right)\left(\overline{\sigma_{k}}\otimes\sigma_{k}-I\right)\nonumber \\
 & = & \gamma\left[\left(\alpha_{1}\left(t\right)+\alpha_{2}\left(t\right)\right)\mathrm{adiag}\left(1,0,0,1\right)+\left(\alpha_{1}\left(t\right)-\alpha_{2}\left(t\right)\right)\mathrm{adiag}\left(0,1,1,0\right)\right.\nonumber \\
 &  & \left.-\left(\alpha_{1}\left(t\right)+\alpha_{2}\left(t\right)\right)I-2\alpha_{3}\left(t\right)\mathrm{diag}\left(0,1,1,0\right)\right],\label{eq:2level_generator_matrix_form}
\end{eqnarray}
where $I$ is the identity matrix, 
\[
\mathrm{adiag}\left(a_{1},\ldots,a_{n}\right)=\left(\begin{array}{ccc}
O &  & a_{1}\\
 & \iddots\\
a_{n} &  & O
\end{array}\right),
\]
and 
\[
\mathrm{vec}\left(\begin{array}{ccc}
a_{11} & \cdots & a_{1n}\\
\vdots & \ddots & \vdots\\
a_{n1} & \cdots & a_{nn}
\end{array}\right)=\left(\begin{array}{c}
a_{11}\\
\vdots\\
a_{n1}\\
a_{12}\\
\vdots\\
a_{n2}\\
\vdots\\
a_{nn}
\end{array}\right)
\]
is the vector form of the matrix, which is used to represent a superoperator
in its matrix form. One can easily check $L$ is commutative by a
direct calculation. 

In (\ref{eq:2level_generator_matrix_form}), put
\[
\begin{array}{cc}
G_{1}=\mathrm{adiag}\left(1,0,0,1\right), & f_{1}\left(t,\lambda\right)=\gamma\left\{ \alpha_{1}\left(t\right)+\alpha_{2}\left(t\right)\right\} \lambda,\\
G_{2}=\mathrm{adiag}\left(0,1,1,0,\right) & f_{2}\left(t,\lambda\right)=\gamma\left\{ \alpha_{1}\left(t\right)-\alpha_{2}\left(t\right)\right\} \lambda,\\
G_{3}=I, & f_{3}\left(t,\lambda\right)=-\gamma\left\{ \alpha_{1}\left(t\right)+\alpha_{2}\left(t\right)\right\} \lambda,\\
G_{4}=\mathrm{diag}\left(0,1,1,0\right), & f_{4}=-2\gamma\alpha_{3}\left(t\right)\lambda,
\end{array}
\]
then, applying Theorem \ref{thm:zhu_spatial_decomposition_solution} we have 
{\footnotesize 
\begin{eqnarray*}
g_{1}\left(t,G_{1}\right) & = & C_{1}\sum_{\xi=0}^{\infty}\frac{\left(\int_{0}^{t}\gamma\left\{ \alpha_{1}\left(\tau\right)+\alpha_{2}\left(\tau\right)\right\} d\tau\right)^{\xi}}{\xi!}G_{1}^{\xi}\\
 & = & C_{1}\left[G_{4}+\sum_{\xi=0}^{\infty}\frac{\left(\int_{0}^{t}\gamma\left\{ \alpha_{1}\left(\tau\right)+\alpha_{2}\left(\tau\right)\right\} d\tau\right)^{2\xi+1}}{\left(2\xi+1\right)!}G_{1} \right.\\
&& \left. +\sum_{\xi=0}^{\infty}\frac{\left(\int_{0}^{t}\gamma\left\{ \alpha_{1}\left(\tau\right)+\alpha_{2}\left(\tau\right)\right\} d\tau\right)^{2\xi}}{\left(2\xi\right)!}\mathrm{diag}\left(1,0,0,1\right)\right]\\
 & = & C_{1}\left[G_{4}+\cosh\left(\int_{0}^{t}\gamma\left\{ \alpha_{1}\left(\tau\right)+\alpha_{2}\left(\tau\right)\right\} d\tau\right)\mathrm{diag}\left(1,0,0,1\right)\right.\\
 &  & \left.+\sinh\left(\int_{0}^{t}\gamma\left\{ \alpha_{1}\left(\tau\right)+\alpha_{2}\left(\tau\right)\right\} d\tau\right)\mathrm{adiag}\left(1,0,0,1\right)\right]
\end{eqnarray*}
\begin{eqnarray*}
g_{2}\left(t,G_{2}\right) & = & C_{2}\sum_{\xi=0}^{\infty}\frac{\left(\int_{0}^{t}\gamma\left\{ \alpha_{1}\left(\tau\right)-\alpha_{2}\left(\tau\right)\right\} d\tau\right)}{\xi!}G_{2}^{\xi}\\
 & = & C_{2}\left[ \mathrm{diag}(1,0,0,1)+\sum_{\xi=0}^{\infty}\frac{\left(\int_{0}^{t}\gamma\left\{ \alpha_{1}\left(\tau\right)-\alpha_{2}\left(\tau\right)\right\} d\tau\right)^{2\xi+1}}{\left(2\xi+1\right)!}G_{2} \right. \\
&& \left. +\sum_{\xi=0}^{\infty}\frac{\left(\int_{0}^{t}\gamma\left\{ \alpha_{1}\left(\tau\right)-\alpha_{2}\left(\tau\right)\right\} d\tau\right)^{2\xi}}{\left(2\xi\right)!} G_{4} \right] \\
 & = & C_{2}\left[ \mathrm{diag}(1,0,0,1)+\sinh\left(\int_{0}^{t}\gamma\left\{ \alpha_{1}\left(\tau\right)-\alpha_{2}\left(\tau\right)\right\} d\tau\right)G_{2} \right. \\
&& \left. +\cosh\left(\int_{0}^{t}\gamma\left\{ \alpha_{1}\left(\tau\right)-\alpha_{2}\left(\tau\right)\right\} d\tau\right) G_{4} \right]
\end{eqnarray*}
\begin{eqnarray*}
g_{3}\left(t,G_{3}\right) & = & C_{3}\exp\left(-\gamma\int_{0}^{t}\left\{ \alpha_{1}\left(\tau\right)+\alpha_{2}\left(\tau\right)\right\} d\tau\right)I\\
g_{4}\left(t,G_{4}\right) & = & C_{4} \left[ \mathrm{diag}(1,0,0,1)+ \exp\left(-2\gamma\int_{0}^{t}\alpha_{3}\left(t\right)d\tau\right) G_{4} \right],
\end{eqnarray*}}
where $C_{1},\ldots,C_{4}\in\mathbb{C}$ are arbitrary constants.
Then we obtain the fundamental solution $\Phi\left(t\right)$ of (\ref{eq:master_equation})
{\footnotesize 
\begin{eqnarray*}
\Phi\left(t\right) & = & \prod_{k=1}^{4}g_{k}\left(t,G_{k}\right)\\
 & = & \left. C\exp\left(-\gamma\int_{0}^{t}\left\{ \alpha_{1}\left(\tau\right)+\alpha_{2}\left(\tau\right)\right\} d\tau\right)\right[ \\
&& \cosh \left( \int_{0}^{t}\gamma \left\{ \alpha_{1}(\tau)+\alpha_{2}(\tau) \right\} d\tau \right) \mathrm{diag}(1,0,0,1) + \sinh \left( \int_{0}^{t}\gamma \left\{ \alpha_{1}(\tau)+\alpha_{2}(\tau) \right\} d\tau \right) \mathrm{adiag}(1,0,0,1) \\
&& +\sinh \left( \int_{0}^{t}\gamma \left\{ \alpha_{1}(\tau)-\alpha_{2}(\tau) \right\} d\tau \right) \exp \left( -2\gamma \int_{0}^{t} \alpha_{3}(\tau) d\tau \right) \mathrm{adiag}(0,1,1,0) \\
&& \left. + \cosh \left( \int_{0}^{t}\gamma \left\{ \alpha_{1}(\tau)-\alpha_{2}(\tau) \right\} d\tau \right) \exp \left( -2\gamma \int_{0}^{t} \alpha_{3}(\tau) d\tau \right) \mathrm{diag}(0,1,1,0) \right] \\
&=& \frac{C}{2}\left[ \left( 1+ \exp \left( -2\gamma \int_{0}^{t} \left\{ \alpha_{1}(\tau)+\alpha_{2}(\tau) \right\} d\tau \right) \right) \mathrm{diag}(1,0,0,1) \right. \\
&& +\left( 1- \exp \left( -2\gamma \int_{0}^{t} \left\{ \alpha_{1}(\tau)+\alpha_{2}(\tau) \right\} d\tau \right) \right) \mathrm{adiag}(1,0,0,1) \\
&& +\left( \exp \left( -2\gamma \int_{0}^{t} \alpha_{2}(\tau) d\tau \right) -\exp \left( -2\gamma \int_{0}^{t} \alpha_{1}(\tau) d\tau \right) \right) \exp \left( -2\gamma \int_{0}^{t} \alpha_{3}(\tau) d\tau \right) \mathrm{adiag}(0,1,1,0) \\
&& +\left. \left( \exp \left( -2\gamma \int_{0}^{t} \alpha_{2}(\tau) d\tau \right) +\exp \left( -2\gamma \int_{0}^{t} \alpha_{1}(\tau) d\tau \right) \right) \exp \left( -2\gamma \int_{0}^{t} \alpha_{3}(\tau) d\tau \right) \mathrm{diag}(0,1,1,0) \right] ,
\end{eqnarray*}}
where $C=C_{1}C_{2}C_{3}C_{4}$ is some constant, or in another form
{\footnotesize 
\begin{eqnarray*}
\Phi\left(t\right)\rho & = & \frac{C}{2}\left[ \left( 1+ \exp \left( -2\gamma \int_{0}^{t} \left\{ \alpha_{1}(\tau)+\alpha_{2}(\tau) \right\} d\tau \right) \right) \left( \sigma_{3}\rho\sigma_{3}+\rho \right) \right. \\
&& +\left( 1- \exp \left( -2\gamma \int_{0}^{t} \left\{ \alpha_{1}(\tau)+\alpha_{2}(\tau) \right\} d\tau \right) \right) \left( \sigma_{1} \rho \sigma_{1}+\sigma_{2} \rho \sigma_{2} \right) \\
&& +\left( \exp \left( -2\gamma \int_{0}^{t} \alpha_{2}(\tau) d\tau \right) -\exp \left( -2\gamma \int_{0}^{t} \alpha_{1}(\tau) d\tau \right) \right) \exp \left( -2\gamma \int_{0}^{t} \alpha_{3}(\tau) d\tau \right) \left( \sigma_{1} \rho \sigma_{1} - \sigma_{2} \rho \sigma_{2} \right) \\
&& +\left. \left( \exp \left( -2\gamma \int_{0}^{t} \alpha_{2}(\tau) d\tau \right) +\exp \left( -2\gamma \int_{0}^{t} \alpha_{1}(\tau) d\tau \right) \right) \exp \left( -2\gamma \int_{0}^{t} \alpha_{3}(\tau) d\tau \right) \left( \rho -\sigma_{3} \rho \sigma_{3} \right) \right].
\end{eqnarray*}}
Since $\Phi(0)=I$, we have $C=1$. 
\end{example}

\begin{example}
We consider another 2-level system 
\begin{eqnarray*}
\frac{d}{dt}\rho_{t} & = & L\left(t\right)\rho_{t}\\
L\left(t\right)\rho & = & -\frac{i}{2}\varepsilon\left(t\right)\left[\sigma_{3},\rho\right]+\gamma\left(t\right)\left(\mu L_{1}+\left(1-\mu\right)L_{2}\right)\rho\\
 &  & +\frac{1}{2}\sum_{\alpha,\beta=0}^{1}c_{\alpha,\beta}\left(t\right)\left(\left[F_{\alpha},\rho F_{\beta}\right]+\left[F_{\alpha}\rho,F_{\beta}\right]\right),
\end{eqnarray*}
where $\mu\in\left[0,1\right]$ is some constant, 
\[
\begin{array}{cc}
L_{1}\rho=\sigma^{+}\rho\sigma^{-}-\frac{1}{2}\left\{ \sigma^{-}\sigma^{+},\rho\right\} , & L_{2}\rho=\sigma^{-}\rho\sigma^{+}-\frac{1}{2}\left\{ \sigma^{+}\sigma^{-},\rho\right\} ,\\
F_{0}=\sigma^{-}\sigma^{+}, & F_{1}=\sigma^{+}\sigma^{-},\\
\sigma^{\pm}=\frac{1}{\sqrt{2}}\left(\sigma_{1}\pm i\sigma_{2}\right),
\end{array}
\]
 $\left[A,B\right]=AB-BA$, $\left\{ A,B\right\} =AB+BA$ and $\gamma,\varepsilon,c_{\alpha,\beta}$
are time-dependent complex functions. This example was considered in \cite{chrusinski2010},
but here we calculate the solution using the Zhu's method. By a simple calculation
one finds the matrix form of $L\left(t\right)$ is given by
{\footnotesize 
\begin{eqnarray*}
L & = & \left(\begin{array}{cccc}
2\gamma\left(\mu-1\right) & 0 & 0 & 2\gamma\mu\\
0 & -2c_{00}+4c_{01}-2c_{11}+i\varepsilon-\gamma & 0 & 0\\
0 & 0 & -2c_{00}+4c_{10}-2c_{11}-i\varepsilon-\gamma & 0\\
2\gamma\left(1-\mu\right) & 0 & 0 & -2\gamma\mu
\end{array}\right)\\
 & = & A_{1}\left(t\right)+A_{2}\left(t\right)+A_{3}\left(t\right),
\end{eqnarray*}}
where 
\begin{eqnarray*}
A_{1}\left(t\right) & = & 2\gamma\left(t\right)G_{1}\\
A_{2}\left(t\right) & = & \left\{ 4c_{01}\left(t\right)-2c_{00}\left(t\right)-2c_{11}\left(t\right)-\gamma\left(t\right)+i\varepsilon\left(t\right)\right\} G_{2}\\
A_{3}\left(t\right) & = & \left\{ 4c_{10}\left(t\right)-2c_{00}\left(t\right)-2c_{11}\left(t\right)-\gamma\left(t\right)-i\varepsilon\left(t\right)\right\} G_{3}
\end{eqnarray*}
with
\[ \begin{array}{ccc}
G_{1}  =  \left(\begin{array}{cccc}
\mu-1 & 0 & 0 & \mu\\
0 & 0 & 0 & 0\\
0 & 0 & 0 & 0\\
1-\mu & 0 & 0 & -\mu
\end{array}\right) & G_{2}  =  \mathrm{diag}\left(0,1,0,0\right) & G_{3}  =  \mathrm{diag}\left(0,0,1,0\right) \end{array} \]

\begin{eqnarray*}
f_{1}\left(t,\lambda\right) & = & 2\gamma\left(t\right)\lambda\\
f_{2}\left(t,\lambda\right) & = & \left\{ 4c_{01}\left(t\right)-2c_{00}\left(t\right)-2c_{11}\left(t\right)-\gamma\left(t\right)+i\varepsilon\left(t\right)\right\} \lambda\\
f_{3}\left(t,\lambda\right) & = & \left\{ 4c_{10}\left(t\right)-2c_{00}\left(t\right)-2c_{11}\left(t\right)-\gamma\left(t\right)-i\varepsilon\left(t\right)\right\} \lambda.
\end{eqnarray*}
Thus, by applying Theorem \ref{thm:zhu_spatial_decomposition_solution},
{\footnotesize 
\begin{eqnarray*}
g_{1}\left(t,G_{1}\right) & = & C_{1}\exp\left(\int_{0}^{t}f_{1}\left(\tau,G_{1}\right)d\tau\right)\\
 & = & C_{1} \left[ \left( \begin{array}{cccc} \mu & 0 & 0 & \mu \\ 0 & 1 & 0 & 0 \\ 0 & 0 & 1 & 0 \\ 1-\mu & 0 & 0 & 1-\mu \end{array} \right) + \cosh \left( \int_{0}^{t} 2\gamma(\tau) d\tau \right) \left( \begin{array}{cccc} 1-\mu & 0 & 0 & -\mu \\ 0 & 0 & 0 & 0 \\ 0 & 0 & 0 & 0 \\ \mu-1 & 0 & 0 & \mu \end{array} \right) \right. \\
&& \left. +\sinh  \left( \int_{0}^{t} 2\gamma(\tau) d\tau \right)G_{1} \right] \\
&=& C_{1}\left( \begin{array}{cccc} \mu & 0 & 0 & \mu \\ 0 & 1 & 0 & 0 \\ 0 & 0 & 1 & 0 \\ 1-\mu & 0 & 0 & 1-\mu \end{array} \right) - C_{1}  \exp \left( -\int_{0}^{t} 2\gamma (\tau) d\tau \right) G_{1}
\end{eqnarray*}
\begin{eqnarray*}
g_{2}\left(t,G_{2}\right) & = & C_{2}\exp\left(\int_{0}^{t}f_{2}\left(\tau,G_{2}\right)d\tau\right)\\
 & = & C_{2}\left[I-G_{2}+\sum_{k=0}^{\infty}\frac{\left(\int_{0}^{t}\left\{ 4c_{01}\left(\tau\right)-2c_{00}\left(\tau\right)-2c_{11}\left(\tau\right)-\gamma\left(\tau\right)+i\varepsilon\left(\tau\right)\right\} d\tau\right)^{k}}{k!}G_{2}\right]\\
 & = & C_{2}\mathrm{diag}\left(1,0,1,1\right)+C_{2} \exp\left(\int_{0}^{t}\left\{ 4c_{01}\left(\tau\right)-2c_{00}\left(\tau\right)-2c_{11}\left(\tau\right)-\gamma\left(\tau\right)+i\varepsilon\left(\tau\right)\right\} d\tau\right)G_{2}
\end{eqnarray*}
\begin{eqnarray*}
g_{3}\left(t,G_{3}\right) & = & C_{3}\exp\left(\int_{0}^{t}f_{3}\left(\tau,G_{3}\right)d\tau\right)\\
 & = & C_{3}\left[I-G_{3}+\sum_{k=0}^{\infty}\frac{\left(\int_{0}^{t}\left\{ 4c_{10}\left(\tau\right)-2c_{00}\left(\tau\right)-2c_{11}\left(\tau\right)-\gamma\left(\tau\right)-i\varepsilon\left(\tau\right)\right\} d\tau\right)^{k}}{k!}G_{3}\right]\\
 & = & C_{3}\mathrm{diag}\left(1,1,0,1\right)+C_{3} \exp\left(\int_{0}^{t}\left\{ 4c_{10}\left(\tau\right)-2c_{00}\left(\tau\right)-2c_{11}\left(\tau\right)-\gamma\left(\tau\right)-i\varepsilon\left(\tau\right)\right\} d\tau\right)G_{3},
\end{eqnarray*}}
where $C_{1},C_{2},C_{3}\in\mathbb{C}$ are arbitrary constants. Then,
finally we obtain the solution 
{\footnotesize 
\begin{eqnarray*}
\Phi\left(t\right) & = & C \left(\begin{array}{cccc}
\mu & 0 & 0 & \mu\\
0 & 0 & 0 & 0\\
0 & 0 & 0 & 0\\
1-\mu & 0 & 0 & 1-\mu
\end{array}\right)\\
&& + C \exp\left(\int_{0}^{t}\left\{ 4c_{01}\left(\tau\right)-2c_{00}\left(\tau\right)-2c_{11}\left(\tau\right)-\gamma\left(\tau\right)+i\varepsilon\left(\tau\right)\right\} d\tau\right)G_{2} \\
&& +C \exp\left(\int_{0}^{t}\left\{ 4c_{10}\left(\tau\right)-2c_{00}\left(\tau\right)-2c_{11}\left(\tau\right)-\gamma\left(\tau\right)-i\varepsilon\left(\tau\right)\right\} d\tau\right)G_{3} \\
&& -C \exp \left( -\int_{0}^{t} 2\gamma (\tau) d\tau \right) G_{1}
\end{eqnarray*}}
where $C=C_{1}C_{2}C_{3}$ is a constant, or in another form
{\footnotesize 
\begin{eqnarray*}
\Phi(t)\rho &=& \frac{C}{4} \left[ (2\mu -1)\left( \rho\sigma_{3}+\sigma_{3}\rho +i \sigma_{1}\rho\sigma_{2}+i\sigma_{2}\rho\sigma_{1} \right) +\rho +\sigma_{1}\rho\sigma_{1} -\sigma_{2}\rho\sigma_{2}+\sigma_{3}\rho\sigma_{3} \right] \\
&& +\frac{C}{4} \exp\left(\int_{0}^{t}\left\{ 4c_{01}\left(\tau\right)-2c_{00}\left(\tau\right)-2c_{11}\left(\tau\right)-\gamma\left(\tau\right)+i\varepsilon\left(\tau\right)\right\} d\tau\right) \left[ \rho-\sigma_{3}\rho\sigma_{3}+\rho\sigma_{3}-\sigma_{3}\rho \right] \\
&& +\frac{C}{4} \exp\left(\int_{0}^{t}\left\{ 4c_{10}\left(\tau\right)-2c_{00}\left(\tau\right)-2c_{11}\left(\tau\right)-\gamma\left(\tau\right)-i\varepsilon\left(\tau\right)\right\} d\tau\right) \left[ \rho-\sigma_{3}\rho\sigma_{3}-\rho\sigma_{3}+\sigma_{3}\rho \right] \\
&& -\frac{C}{4} \exp \left( -\int_{0}^{t} 2\gamma (\tau) d\tau \right) \left[ (2\mu -1)\left( \rho\sigma_{3}+\sigma_{3}\rho   +i \sigma_{1}\rho\sigma_{2}+i\sigma_{2}\rho\sigma_{1}  \right) -\rho +\sigma_{1}\rho\sigma_{1}-\sigma_{2}\rho\sigma_{2}-\sigma_{3}\rho\sigma_{3} \right].
\end{eqnarray*}}
The initial condition implies $C=1$. 
\end{example}

\section{Concluding Remark}

In this paper some results on ordinary linear differential equations with functionally commutative generators were introduced, and using the Zhu's result, a method to solve a certain
type of differential equations effectively, some quantum
master equations with functionally commutative generators were solved effectively, i.e. involving only finitely many terms. 
Indeed, this method does not rely on the spectrum of the generator
explicitly, in other words we do not need to calculate the eigenvalues.
This is an advantage from a computational point, and we may be able
to solve higher dimensional and more complicated master equations, where it is difficult to calculate the eigenvalues.

\section*{Acknowledgement}
On this research, I would like to express my appreciation to Professor Andrzej Jamio{\l}kowski and Professor Dariusz Chru{\'s}ci{\'n}ski for valuable advices.

\end{document}